\documentclass[a4paper,aps,prd,twocolumn,preprintnumbers,showpacs,superscriptaddress,nofootinbib]{revtex4}
\usepackage{graphics,graphicx,epsfig}
\usepackage{amsfonts,amsmath,amssymb}
\usepackage{amsmath}
\usepackage{upgreek}
\usepackage{calrsfs} 
\DeclareMathAlphabet{\pazocal}{OMS}{zplm}{m}{n}
\usepackage{amsfonts}
\usepackage{amssymb}
\usepackage{color}
\usepackage{graphicx}
\definecolor{darkgreen}{rgb}{0,0.35,0}

\begin{document}

\title{Torsion as a Gauge Field in a Lorentz-Chern-Simons Theory}
\author{Sim\'on del Pino}
\email{simon.delpino.m-at-mail.pucv.cl}
\affiliation{Instituto de F\'{\i}sica, Pontificia Universidad Cat\'{o}lica de Valpara%
\'{\i}so, Casilla 4059, Valpara\'{\i}so, Chile.}

\author{Adolfo Toloza}
\email{atoloza-at-cecs.cl}
\affiliation{Instituto de F\'{\i}sica, Pontificia Universidad Cat\'{o}lica de Valpara%
\'{\i}so, Casilla 4059, Valpara\'{\i}so, Chile.}
\affiliation{Centro de Estudios Cient\'{\i}ficos CECs Casilla 1469, Valdivia, Chile}
\begin{abstract}
We explore a model of gravity that arises from the consideration of the Chern-Simons form in 2+1 dimensions for a spin connection with a contorsion described by a scalar and a vector field.
The effective Lagrangian presents a local Weyl symmetry allowing us to gauge the scalar field to a constant value. From a gauge field theory perspective, it is shown that the vector part of the torsion (related to its trace) is a gauge field for the Weyl group, which allows the interpretation of the torsion as an electromagnetic field.
In the gauge of constant scalar field we obtain Chiral Gravity coupled to a Chern-Simons-Proca theory for the vector field, that at the level of equations of motion is equivalent to Topologically Massive Electrodynamics minimally coupled to Chiral Gravity. 
Electrodynamics and gravity appear here unified as geometrical features of a Riemann-Cartan manifold.
\end{abstract}

\maketitle

\section{Introduction}   
Dark matter and dark energy represent the majority of content in the Universe. Its nature has led physicists to consider extended gravity frameworks different from original Einstein's general relativity. Among possible extensions, the literature reports that the presence of torsion, in certain gravity theories, display features similar to those of the presently observed accelerating universe compatible with the dark energy hypothesis \cite{pgt,AZ,Magueijo-Zlosnik-Kibble,Tilquin-Schucker}.\\
In the case of lower dimensional scenarios, it has been shown that the presence of torsion can also be seen as the action of effective matter fields \cite{blago}. 
$(2+1)$-dimensional spacetime give rise to simpler models that share the conceptual foundations of four dimensional theories, while avoiding some of the computational difficulties. The systematic studies of lower dimensional models with torsion rely on the fact that it is expected to improve our comprehension of the dynamical role of torsion in gravity. It started with the Mielke-Baekler model \cite{MB} as an extension of topologically massive gravity.

Einstein's gravity is trivial in 2+1 dimensions on the grounds that it does not present propagating degrees of freedom. However, in the case of negative cosmological constant there are asymptotically $AdS_3$ black hole solutions \cite{BTZ}, that resemble all the characteristic features of real black hole solutions for four dimensional theories. Even though the theory from which originally these solutions arose is still considered too simple, they have been proven to provide a good tool to understand the physics of these entities in a very simple model. Topologically massive gauge theories \cite{TMG}, on the other hand, are built with the usual gauge invariant Lagrangian, augmented by a term of topological origin related to the Chern-Simons (CS) secondary characteristic classes \cite{Chern} that exists only in odd dimensions. This term adds a propagating massive degree of freedom in the case of second-rank tensor fields describing gravity, in which case the model is referred to as Topologically Massive Gravity (TMG) whereas that for the $U(1)$ gauge group as Topologically Massive Electrodynamics (TME) \cite{TMG,TMGT}.\\
For a particular choice of coupling constants of TMG, namely when the graviton mass, $\mu$, equals the  inverse of the curvature radius of $AdS_3$, $\mu l=1$, the theory looses the degree of freedom added by the Chern-Simons term, and the action can be written as a Chern-Simons action for a gauge field $\mathcal{A}^{ab}$ \cite{TMG-CG}, reminiscent of the famous Ach\'ucaro-Towsend-Witten factorization of the Einstein gravity \cite{ATW}. In \cite{CG}, Li, Song and Strominger argued inconsistency of TMG for generic coupling constants except for this particular (chiral) point of the parameters space, known as Chiral Gravity (CG). Provided boundary conditions, CG may represent a consistent quantum theory of gravity for $AdS_3$ spacetimes \cite{MSS}, although some question are still open, among others, how to generalize CG to couple matter fields and local degrees of freedom. 
It was shown in \cite{US} that CG can be thought as a Lagrangian for a spin connection of a Riemann-Cartan (RC) manifold of constant completely-antisymmetric torsion. The inverse of $AdS_3$ radius, $l^{-1}$, in this work was promoted to a scalar field, adding a degree of freedom to the theory, that when acquires a constant value reproduces the gauge connection $\mathcal{A}^{ab}$ (see section \ref{CG&LCS}). The extension to that model motivates this investigation. The question of how  to incorporate more torsional degrees of freedom in the picture, in order to understand more fundamentally its behaviour and what kind of couplings will arise in this procedure is here addressed. We include here the trace of the torsion $2-$form encoded in the $1-$form,  $A\sim\iota_aT^a$\footnote{$\iota_a$ is the contraction operator defined in section IV.}.\\
The paper is organized as follows: Section \ref{CG&LCS} motivates our model by explaining the Chern-Simons decomposition of TMG and how we add torsional degrees of freedom from it. It also shows the effective theory we will work with. Section \ref{Weyl symmetry} is dedicated to the Weyl symmetry present in the model, and explains how the trace of torsion can be understood as a gauge field for this symmetry. The details of the equations of motion and the relation of our model with TME is explained in section \ref{dyn}. Conclusions and remarks are given next, in section \ref{conclusion}.

\section{Chiral Gravity and Lorentz-Chern-Simons Theory}\label{CG&LCS}   
Let $\tilde{\omega}^{ab}$ be a Riemannian  spin connection 1-form that describes the affine structure of a three dimensional manifold. Let $e^a=e^{a}_{\ \mu}dx^\mu$ be the dreibein 1-form that defines the metric structure of the same manifold by the relation $g_{\mu\nu}=e^{a}_{\ \mu}e^{b}_{\ \nu}\eta_{ab}$. The dreibein is the mapping that relates Greek characters (coordinate indices) with the Latin ones (Lorentz indices). In our convention the flat Lorentzian metric will be $\eta_{ab}=\,$diag$\,(-,+,+)$. The cosmological TMG action accepts the following factorization for the gauge connections $\mathcal{A}^{ab}_{\pm}=\tilde{\omega}^{ab}\pm\frac{1}{l}\epsilon^{ab}_{\ \ c}e^c$ \cite{TMG-CG}.
\begin{equation}\label{TMG}
I_{TMG}=-\frac{1}{2}\left(1-\frac{1}{\mu l}\right)I_{CS}[\mathcal{A}_{+}]+\frac{1}{2}\left(1+\frac{1}{\mu l}\right)I_{CS}[\mathcal{A}_{-}],
\end{equation}
expressed in terms of the CS action
\begin{equation*}
I_{CS}[\mathcal{A}]=\frac{1}{2}\int\limits_{M_3}\left(\mathcal{A}^a{}_b \wedge d\mathcal{A}^b{}_a + \frac{2}{3}\mathcal{A}^a{}_b \wedge \mathcal{A}^b{}_c \wedge \mathcal{A}^c{}_a\right).
\end{equation*}
Note that in action (\ref{TMG}) the dreibein (or equivalently the metric $g_{\mu\nu}$) is the only independent field, since $e^a$ and $\tilde{\omega}^{ab}$ are related through the second Cartan's structure equation that states the vanishing of the torsion made out of $\tilde{\omega}$,
\begin{equation}\label{torsionless}
\tilde{T}^a=\tilde{D}e^a=de^a+\tilde{\omega}^{a}_{\ b}\wedge e^b=0,
\end{equation} 
fixing completely the components of $\tilde{\omega}$ in terms of the dreibein and its first derivatives \cite{Zanelli}. 
The particular choice of parameters, 
\begin{equation*}
\mu l=1
\end{equation*}
defines the so called chiral point of TMG (CG) which makes one of the CS copies in (\ref{TMG}) to vanish, making the model to be a single CS Lagranian for a gauge connection $\mathcal{A}^{ab}_{-}=\tilde{\omega}^{ab}-\frac{1}{l}\epsilon^{ab}_{\ \ c}e^c$ that describes the dynamics of a three dimensional Riemannian manifold.\\
A RC geometry allows the presence of torsion in the manifold, and hence it does not relate the spin connection and the dreibein through a relation such as (\ref{torsionless}). Here we want to generalize the action of CG to a RC geometry. To describe a manifold with torsion it is necessary to allow its affine structure to enter as a dynamical field in the same footing as the dreibein does for its metric description. We call this spin connection $\omega^{ab}$, and it incorporates torsional degrees of freedom in the model. It is also related to the Riemannian one by simply
\begin{equation}
\omega ^{ab}=\tilde{\omega}^{ab}+\kappa ^{ab},  \label{SpinConnection}
\end{equation}
where $\kappa^{ab}$ is the contorsion 1-form, which carries the torsional degrees of freedom. In fact we see from here that, since $\tilde{\omega}$ still satisfies (\ref{torsionless}),
\begin{equation*}
T^a=De^a=\kappa^a_{\ b}\wedge e^b,
\end{equation*}
which is non zero in general. There is an irreducible decomposition to write the contorsion in terms of its trace and axial parts as
\begin{equation}\label{contorsion decomp}
\kappa^{ab}=-\phi\epsilon^{ab}_{\ \ c}e^c-A^{[a}e^{b]}+M^{ab},
\end{equation}
where $\phi$ is a (pseudo-)scalar field for the completely antisymmetric, or axial part, $A^a$ the vector part, related to the trace of the torsion, and $M^{ab}=M^{ab}_{\ \ c} e^c$ is the mixed part which is constrained to have vanishing trace and completely antisymmetric part ($M_{[abc]}=0$, $M^{ab}_{\ \ b}=0$). Here $X_{[a_1a_2...a_n]}=\frac{1}{n!}(X_{a_1a_2...a_n}\pm \text{permutations})$ denotes normalized anti-symmetrization. It is clear that the spin connection $\omega^{ab}$ becomes $\mathcal{A}^{ab}_{-}$ when it has an axial contorsion 1-form of constant value parametrized by $l^{-1}$. Hence CG can be interpreted as a CS theory for a connection with constant antisymmetric torsion \cite{US}.\\ 
Consider now the three-dimensional CS Lagrangian for a general spin connection, 
\begin{equation}
\mathcal{L}_{CS}(\omega )=\omega^a{}_b \wedge d\omega^b{}_a + \frac{2}{3}\omega^a{}_b \wedge \omega^b{}_c \wedge \omega^c{}_a . \label{CSLagrangian}
\end{equation}
Now use (\ref{SpinConnection}) as a general ansatz. In $\mathcal{L}_{CS}$ it yields
\begin{align}\label{splitted lagrangian}
\mathcal{L}_{CS}(\omega) &  =\mathcal{L}_{CS}(\tilde{\omega})+2\kappa^a{}_b \wedge\tilde{R}^b{}_a  +\frac{2}{3}\kappa^a{}_b \wedge \kappa^b{}_c \wedge\kappa^c{}_a  \nonumber\\
&  +\kappa^a{}_b \wedge\tilde{D} \kappa^b{}_a + d \left(\tilde{\omega}^a{}_b \wedge\kappa^b{}_a \right),
\end{align}
where $\tilde{R}=d\tilde{\omega}+\tilde{\omega}\wedge\tilde{\omega}$ is the Riemannian curvature. The minimal model incorporates $\phi$ as the single new dynamical field and it was studied in detail in \cite{US} leading to an effective theory that couples the scalar field non minimally to geometry. We call $\mathcal{L}_{CS}^{ax}$ to Lagrangian \eqref{splitted lagrangian} with purely axial contorsion in \eqref{contorsion decomp},
\begin{eqnarray}
I^{ax}_{CS}&\equiv &\frac{k}{4\pi }\int\limits_{M_{3}}\mathcal{L}_{CS}^{ax}(\omega )\notag\\
&=& \frac{k}{2\pi }\int\limits_{M_{3}}\bigg[\phi \epsilon _{abc}\left( \tilde{R}^{ab}+\frac{1}{3}\phi ^{2}e^{a}\wedge e^{b}\right) \wedge e^{c}+\frac{1}{2}\mathcal{L}_{CS}(\tilde{\omega})\notag\\
&&+\frac{1}{2}d\left( \phi \epsilon _{abc}\tilde{\omega}^{ab}\wedge e^{c}\right) \bigg].  \label{S}
\end{eqnarray}
Here $k$ is the level of the Chern-Simons action, and is given by a positive integer number. We work in units in which the action is dimensionless. Action (\ref{S}) is a particular version of TMG non minimally coupled to a scalar field at the chiral point, invariant under global Weyl transformations 
\begin{equation}\label{WWW}
\phi\rightarrow e^{-\Omega}\phi,\quad e^a\rightarrow e^{\Omega}e^a,
\end{equation}
where $\Omega$ is a constant parameter.
It is important to keep in mind that an overall tilde ( $\tilde{}$ ) denotes quantities built with the Riemannian connection $\tilde{\omega}$, which was previously defined to satisfy (\ref{torsionless}) $\tilde{T}=0$. This constraint is implemented by adding to (\ref{S}) a Lagrange multiplier \cite{Olivera}
\begin{equation}
\frac{k}{4\pi }\int\limits_{M_{3}}\zeta _{a}\wedge \tilde{T}^{a},  \label{Lm}
\end{equation}
where $\zeta _{a}$ is a vector-valued 1-form. This allows us to vary $e^a$ and $\tilde{\omega}^{ab}$ independently preserving condition (\ref{torsionless}).\\
Here we address the next extension of the model by including the vector part of the contorsion,
\begin{equation}\label{effective spin}
\omega^{ab}=\tilde{\omega}^{ab} {-} \phi \epsilon^{ab}{}_c e^c-A^{[a}e^{b]}.
\end{equation} 
This incorporates four out of the nine independent components of $\kappa ^{ab}$.
In (\ref{splitted lagrangian}) it yields an action for $e^a$, $\phi$ and the 1-form $A\equiv A_\mu dx^\mu$
\begin{align}
\mathcal{L}_{CS}^{ax-vec} &  =2\phi\epsilon
_{abc}\left(\tilde{R}^{ab} + \frac{1}{3}\phi^{2}e^a\wedge e^b\right)\wedge e^{c}+\mathcal{L}_{CS}(\tilde{\omega})\nonumber \\
& -\phi A\wedge\ast A - \frac{1}{2}A\wedge dA +4d\phi\wedge\ast A \nonumber\\
& + d\left(\phi\epsilon_{abc}\tilde{\omega}^{ab}\wedge e^{c}+A_{a}\tilde{\omega}^{ab}\wedge e_{b}-2\phi\ast A\right),\label{Full-Lagragian}
\end{align}
where $\ast$ stands for the Hodge dual defined to act on a $p-$form as
$$\ast p = \frac{1}{\left( 3-p\right)!\,p!}\epsilon^{a_1 \cdots a_p}{}_{a_{p+1}\cdots a_3} p_{a_1...a_p}e^{a_{p+1}}\wedge \cdots \wedge e^{a_3}.$$

\section{Local Weyl Symmetry}\label{Weyl symmetry}
Weyl transformations where studied in \cite{Obu,Nieh,Frolov} in the context of Riemann-Cartan spacetimes. Here we show that this symmetry is also present in Lagrangian (\ref{Full-Lagragian}) and that the former can be regarded as the Weyl localized version of (\ref{S}). 
\subsection{Weyl Symmetry of a CS Lagrangian}
Lagrangian (\ref{CSLagrangian}) is purely topological and it does not depend explicitly on the dreibein. Due to this fact it posses a hidden symmetry through its transformation. A Weyl rescaling such as 
\begin{eqnarray}\label{WeylTrans1}
e^a&\rightarrow& e^{\Omega}e^a,\\
\label{WeylTrans2}
\omega^{ab}&\rightarrow&\omega^{ab},
\end{eqnarray}
where $\Omega=\Omega(x)$ is a point dependent function, leaves the action (\ref{CSLagrangian}) invariant. A contorsion with antisymmetric part only, does not posses the amount of fields required to compensate the transformation induced in $\tilde{\omega}$ due to (\ref{WeylTrans1}), in order to preserve the sum $\omega=\tilde{\omega}+\kappa$ invariant. That is the reason why this symmetry was not present in \cite{US}. Here, on the contrary, the symmetry remains after the decomposition (\ref{effective spin}) that led us to (\ref{Full-Lagragian}). Indeed, a transformation such as (\ref{WeylTrans1}) will change $\tilde{\omega}$ by imposing that (\ref{torsionless}) to hold in the new frame, according to
\begin{equation*}
\tilde{\omega}^{ab}\rightarrow \tilde{\omega}^{ab} - 2 \partial^{[a}\Omega e^{b]},
\end{equation*}
thus $\kappa$ must change as
\begin{equation*}
\kappa^{ab}\rightarrow\kappa^{ab}+2\partial^{[a}\Omega e^{b]},
\end{equation*}
in order for (\ref{WeylTrans2}) to hold, regarding (\ref{SpinConnection}). We find that the effective transformation at the level of scalar and vector field that defines this contorsion are
\begin{equation}
A\rightarrow A - 2d\Omega \ \ \ \ , \ \ \ \ \phi\rightarrow e^{-\Omega}\phi.\label{Conf-Trans}
\end{equation}  
Lagrangian (\ref{Full-Lagragian}) is invariant under (\ref{WeylTrans1}) and (\ref{Conf-Trans}).

\subsection{The Weyl-Gauged CG Theory}
From a gauge theory point of view, transformation (\ref{Conf-Trans}) suggests $A$ as a gauge connection for a Weyl symmetry. In fact we will demonstrate here that Lagrangian (\ref{Full-Lagragian}) is the gauge invariant version of (\ref{S}) under Weyl rescalings, to which $A$ is its compensating field.
A general procedure for Weyl-gauging TMG was made in \cite{weyltmg}.\\
Take action \eqref{S}, and perform the usual procedure of localization of its global symmetry \eqref{WWW}, when the transformation parameter is promoted to a local function $\Omega\rightarrow\Omega(x)$, and introduce a gauge field $A$. This field defines a covariant exterior derivative for the Weyl group, symbolically $\mathcal{D}=d-\frac{1}{2}A$, in such a way that it makes the derivatives to transform covariantly under Weyl rescalings,
\begin{eqnarray*}
\mathcal{D}^\prime\phi^\prime=e^{-\Omega}\mathcal{D}\phi\quad&,&\quad\mathcal{D}\phi=d\phi-\frac{1}{2}A\phi,\\
\mathcal{D}^\prime e^{a\prime}=e^{\Omega}\mathcal{D}e^{a}\quad&,&\quad\mathcal{D}e^{a}=de^a+\frac{1}{2}A\wedge e^a,
\end{eqnarray*}
enforcing $A$ to transform as in (\ref{Conf-Trans}). Next we will write action (\ref{S}) in terms of covariant derivatives instead of partial ones,
\begin{equation*}
\mathcal{L}^{ax}(e,\partial e;\phi,\partial\phi)\longrightarrow\mathcal{L}^{ax}(e,\mathcal{D}e;\phi,\mathcal{D}\phi;A).
\end{equation*} 
To do this let us write the Weyl-covariant version of the second Cartan's structure equation (\ref{torsionless}),
\begin{equation*}
\mathcal{D}e^a+\tilde{w}^a_{\ b}\wedge e^b=0,
\end{equation*}
where $\tilde{w}^{ab}$ is invariant. From this connection we construct a Weyl-invariant curvature,
\begin{equation*}
\tilde{\mathcal{R}}^{ab}=d\tilde{w}^{ab}+\tilde{w}^a_{\ c}\wedge\tilde{w}^{cb}.
\end{equation*}
Thus, the local Weyl-invariant version of action (\ref{S}) is made out of these new curvature and connection, 
\begin{eqnarray}\label{action invariant}
\frac{1}{2}\mathcal{L}^{ax}_{\text{W-Inv.}}&=&\phi\epsilon_{abc}\left(\tilde{\mathcal{R}}^{ab}+\frac{1}{3}\phi^2e^a\wedge e^b\right)\wedge e^c\nonumber\\
&&+\frac{1}{2}\mathcal{L}_{CS}(\tilde{w})+\frac{1}{2}d\left(\phi\epsilon_{abc}\tilde{w}^{ab}\wedge e^c\right).
\end{eqnarray}
Note that 
\begin{eqnarray}\label{weyl spin}
\tilde{w}^{ab}&=&\tilde{\omega}^{ab}-A^{[a}e^{b]},\\
\label{weyl curvature}
\tilde{\mathcal{R}}^{ab}&=&\tilde{R}^{ab}+\tilde{D}A^{[a}\wedge e^{b]}+\frac{1}{2}A\wedge A^{[a}e^{b]}\nonumber\\
&&-\frac{1}{4}A^2e^a\wedge e^b,
\end{eqnarray}
where $A^2=A_aA^a$. Finally, replacing (\ref{weyl spin},\ref{weyl curvature}) in (\ref{action invariant}) yields to (\ref{Full-Lagragian}) that concludes the demonstration.\\
Thus the inclusion of the trace of torsion in (\ref{CSLagrangian}) is equivalent to the localization procedure used in gauge theories to give rise to interactions. This enables the interpretation of $\phi$ and $A$ as matter fields.


\section{Dynamics}\label{dyn}
Varying the action with respect to $\phi$, $A$ and $e$ gives the equations of motion that determine our dynamics. In order to achieve that, we have to incorporate the constraint (\ref{Lm}) to (\ref{Full-Lagragian}) and then vary also with respect to $\tilde{\omega}$ and $\zeta$. The action is given by
\begin{equation*}
I^{ax-vec}_{CS}=\frac{k}{2\pi}\int\limits_{M_3}\left(\frac{1}{2}\mathcal{L}^{ax-vec}_{CS}+\frac{1}{2}\zeta^a\wedge\tilde{T}_a\right).
\end{equation*}
The Lagrange multiplier will naturally appear in the equations of motion, so the procedure ends when $\zeta^a$ has been taken out of the system \cite{US}. Due to the Weyl symmetry, the scalar field can be gauged to a constant $l^{-1}$. In that case (\ref{Full-Lagragian}) is a Chern-Simons-Proca (CSP) theory for $A$, minimally coupled to CG. 
\begin{equation*}
I^{ax-vec}_{CS}=I_{CG}+I_{CSP},
\end{equation*}
where
\begin{eqnarray*}
I_{CG}&=&\frac{\bar{k}}{2\pi}\int\limits_{M_3}\bigg[\epsilon
_{abc}\left(\tilde{R}^{ab}+ \frac{1}{3l^{2}}e^a\wedge e^b\right)\wedge e^{c}\nonumber\\
&&+\frac{l}{2}\mathcal{L}_{CS}(\tilde{\omega})+\frac{1}{2}\zeta^a\wedge\tilde{T}_a \bigg],\\
I_{CSP}&=&-\frac{\bar{k}}{4\pi}\int\limits_{M_3}\left(A\wedge\ast A + \frac{l}{2}A\wedge dA\right),
\end{eqnarray*}
and $\bar{k}=\frac{k}{l}$. Boundary terms in this Lagrangian were omitted since they are irrelevant for this discussion. After eliminating $\zeta$, the equations of motion are 
\begin{eqnarray}
\label{delta A}
\delta A&:&\ast dA-\frac{2}{l} A=0,\\ 
\delta e^c&:&\frac{1}{2}\epsilon_{abc}\left(\tilde{R}^{ab}+\frac{1}{l^2}e^a\wedge e^b\right)-lC_c=\frac{1}{2}\tau_c^{CSP}, \label{delta e}
\end{eqnarray}
where 
\begin{equation}
\tau_c^{CSP}=-\frac{1}{2}\left[A\,\iota_c\ast A+\iota_cA\ast A\right], \label{stressCSP}
\end{equation}
is the stress 2-form of the CSP field, and $C_a$ is the Cotton 2-form defined as 
\begin{equation*}
C^a=\tilde{D}\left(\tilde{R}^a-\frac{1}{4}e^a\tilde{R}\right),
\end{equation*}
with
\begin{equation*}
R^a=\iota_{b}\tilde{R}^{ab}\quad,\quad\tilde{R}=\iota_a\tilde{R}^a.
\end{equation*}
Here $\iota_a$ is the contraction operator, defined to act on a $p$-form as $\iota_a p=\frac{1}{(p-1)!}p_{aa_1\cdots a_{p-1}}e^{a_1}\wedge\cdots\wedge e^{a_{p-1}}$.

Something rather remarkable is found in this model, namely Eqs. (\ref{delta A},\ref{delta e}) are equivalent to those of Topologically Massive Electrodynamics (TME) \cite{TMG,TMGT} minimally coupled to CG, up to a gauge fixing.

To show this, let us consider TME minimally coupled to CG
\begin{equation*}
\bar{I}=I_{CG}+I_{TME}
\end{equation*}
where
\begin{eqnarray*}
I_{TME}&=&\frac{\bar{k}}{4\pi}\int\limits_{M_3}\frac{l^2}{4}\left(-F\wedge\ast F+\nu A\wedge F\right).
\end{eqnarray*}
Here, $F=dA$ is the strength form of the gauge field and $\nu$ is a coupling parameter with units of mass. In order to fix the dimensions of the action, a dimensional factor must be included. Without loss of generality it was set to be $l^2/4$ for convenience.\\
The equations of motion are
\begin{eqnarray}
\label{delta A2}
\delta A&:&d\ast F-\nu F=0,\\ 
\label{delta e2}
\delta e^c&:&\frac{1}{2}\epsilon_{abc}\left(\tilde{R}^{ab}+\frac{1}{l^2}e^a\wedge e^b\right)-lC_c=\frac{1}{2}\tau_d^{TME}, 
\end{eqnarray}
where 
\begin{equation}
\tau_c^{TME}=-\frac{l^2}{8}\left[\iota_c(dA)\wedge\ast (dA)-dA\wedge\iota_c\ast dA\right].\label{stressTME}
\end{equation}
is the stress 2-form of the TME field. Eq. (\ref{delta A2}) has a first integral
\begin{equation*}
\ast dA-\nu A=d\Lambda,
\end{equation*}
with $\Lambda$ an integration function, which can be dropped out via a gauge transformation $A\rightarrow A+\frac{1}{\nu}d\Lambda$. Eq. (\ref{delta A}) is equivalent to (\ref{delta A2}) up to a gauge fixing, when the photon mass $\nu$ equals twice the graviton mass, 
\begin{equation}\label{ecuacion de adolfo}
\nu=2\mu=\frac{2}{l}.
\end{equation}
Moreover, we can observe that the stress 2-forms of CSP and TME are equal on-shell
\begin{equation*}
\tau_c^{CSP}\Big|_{A=-\frac{1}{\nu}\ast dA}=\tau_c^{TME}.
\end{equation*}
This means that any solution of CG coupled to TME can be gauged, by a $U(1)$ transformation, to a solution of the system (\ref{delta A},\ref{delta e}) whenever the photon mass has the particular value (\ref{ecuacion de adolfo}), see for instance \cite{CGTMEsolutions}. In consequence, for this model, the 1-form $A$ gravitates as a Maxwell-Chern-Simons field,
resembling the physics of a well known theory of electrodynamics, namely TME.\\
This enforces the interpretation of $A$ as an electromagnetic field. However we must recall that $A$ is not a gauge field for $U(1)$ but for Weyl. This can be seen in the fact that $\phi$ is a real scalar field that carries a current $\mathcal{J}\equiv\delta\mathcal{L}_\phi/\delta A$, where $\mathcal{L}_\phi$ are the terms of the Lagrangian which depend on $\phi$,
\begin{equation*}
\mathcal{J}=\ast\mathcal{D}\phi,
\end{equation*}
which is not an electric current, but a Weyl one. We call the attention to this point on the fact that the equivalence between (\ref{Full-Lagragian}) and TME is manifest in particular gauge fixings of both theories.

\section{Discussion and Conclussions}\label{conclusion}
Matter fields are the origin of curvature of spacetime in gravity theories, where the interaction is expressed in terms of the geometry, and matter content in terms of the energy-momentum tensor. 
A geometric interpretation of the electromagnetic field was attempted by Weyl \cite{Weyl} in a generalization of Riemannian geometry. There, the change of the longitude of a vector when it moves from point to point is compensated by a vector field that enters in the connection of the manifold, and it was identified as the vector potential. Even though did not gain general acceptance, the idea led to the modern gauge principle, corner stone for the understanding of interactions. Dirac returned to Weyl's original ideas in \cite{Dirac}. With the introduction of a gauge (scalar) function, he was able to write down an action invariant under both, Weyl transformations and general coordinate transformations, at the expenses of a transformation of the scalar function. Thus, Dirac obtained a generalized theory for the gravitational field and the electromagnetic field, that in the gauge of constant scalar, results in the known Einstein-Maxwell theory.\\ 
In our model, both the scalar field $\phi$ and the $1-$form $A$ are genuine torsional degrees of freedom, namely its axial part and its trace, and they couple in a CS theory as a Dirac-like scalar and a Weyl-like vector respectively.\\
We have shown a general procedure to introduce torsional degrees of freedom in a gravity action, and we have applied it in a Lorentz-Chern-Simons Lagrangian by means of a decomposition of the contorsion $1-$form.
In a particular gauge, we have shown that there is a Chern-Simons interpretation for Chiral Gravity coupled to Chern-Simons-Proca electrodynamics, that, modulo gauge transformations, shares the solutions space of Chiral Gravity coupled to Topologically Massive Electrodynamics with a particular value of the coupling constant; namely, when the photon mass of TME equals twice the $AdS$ radius. The 1-form $A$ built from the trace of the torsion $2-$form, is interpreted as a gauge field for the Weyl group. 
The constant value of the scalar field determines the strength of the gravity force, through Newton's constant, the cosmological term, and the mass of the photon.

\begin{equation*}
\end{equation*}
The authors would like to thank enlightening discussion with G. Giribet, J. Zanelli, M. Blagojevi\'{c}, O. Mi\v{s}covi\'{c}, R. Olea,  G. Thompson, E. Ay\'on-Beato and A. Anabal\'on.
The work of S.dP. was supported by program MECESUP 0806 and MECESUP CD FSM1204. The Centro de Estudios Cient\'ificos (CECs) is funded by the Chilean Government through the Centers of Excellence Base Financing Program of Conicyt.

\end{document}